\documentclass{elsart}
\usepackage[dvips]{graphicx}
\usepackage{amsmath}
\usepackage{subfigure}
\usepackage{setspace}
\usepackage{lineno}

\begin{document}
\begin{frontmatter}

\title{A New Simple Method for the Analysis of Extensive Air Showers}

\author{H.~Hedayati~Kh.,}
\author{A.~Anvari,}
\author{M.~Bahmanabadi,}
\author{J.~Samimi,}

\address{Department of Physics, Sharif University of Technology,
P.O.Box 11155-9161, Tehran, Iran}

\author{M.~Khakian~Ghomi,}

\address{Department
of Physics, Amirkabir University of Technology, P.O.Box
 15875-4413, Tehran, Iran}

\begin{abstract}
The most important goal of studying an extensive air shower is to
find the energy, mass and arrival direction of its primary cosmic
ray. In order to find these parameters, the core position and
arrival direction of the shower should be determined. In this paper,
a new method for finding core location has been introduced that
utilizes trigger time information of particle detectors. We have
also developed a simple technique to reconstruct the arrival
direction. Our method is not based upon density-sensitive detectors
which are sensitive to the number of crossing particles and is also
independent of lateral distribution models. This model has been
developed and examined by the analysis of simulated shower events
generated by the CORSIKA package.

\end{abstract}


\end{frontmatter}

%\begin{spacing}{2.5}
\setpagewiselinenumbers \modulolinenumbers[1]
\def\linenumberfont{\normalfont\small\sffamily}
\linenumbers
\section{Introduction}
An Extensive Air Shower (EAS) is a large number of secondary
particles originating from a high energy primary cosmic particle. A
lot of efforts have been made to understand the structure and
development of EASs. For an accurate investigation of the EAS
events, we first have to know their direction and core position. The
better the measurement of these parameters, the more precise the
exploration of extensive air shower structures. Plane Wave Front
Approximation (PWFA) is the simplest way to find the direction of
EAS, and spherical front approximation \cite{Gonzalez} is a
complementary approach to obtain more accurate results. The common
method for finding core locations of extensive air showers is to fit
a Lateral Density Function (LDF) to the density of secondary
particles of the shower. Greisen function \cite{Greisen} and its
modified form (e.g. \cite{Nagano}) are usually used as LDF. In LDF
methods, one uses the particle density information without
considering their arrival time information.\\
Some attempts have been made to use mean arrival time and disk
thickness as a measure of the shower core distance \cite{Linsley}.
Measurements of the EAS disk structure were tried by Bassi, Clark,
and Rossi in 1953 \cite{Bassi} and continued later \cite{Agnetta}.
However, it has recently been shown that both mean arrival time and
EAS front thickness in individual showers fluctuate strongly and
cannot be a good measure of the distance from the EAS axis
\cite{Ambrosio}.\\
Though, due to the strong fluctuations, arrival time cannot be used
as a measure of the distance from the core, it can be used to
provide a statistical analysis for some of the EAS features. In this
paper, we propose a model independent approach for finding core
location and arrival direction of EASs which uses arrival time
information, avoids using LDF and can be used for arrays lacking
Density Sensitive Detectors (DSDs), which are sensitive to density
or to the number of crossing particles. In order to examine the
capability of the method, we used the CORSIKA \cite{Heck} package.
\section{Physical Principles}
\begin{figure}
\begin{center}
\includegraphics*[width=1\textwidth]{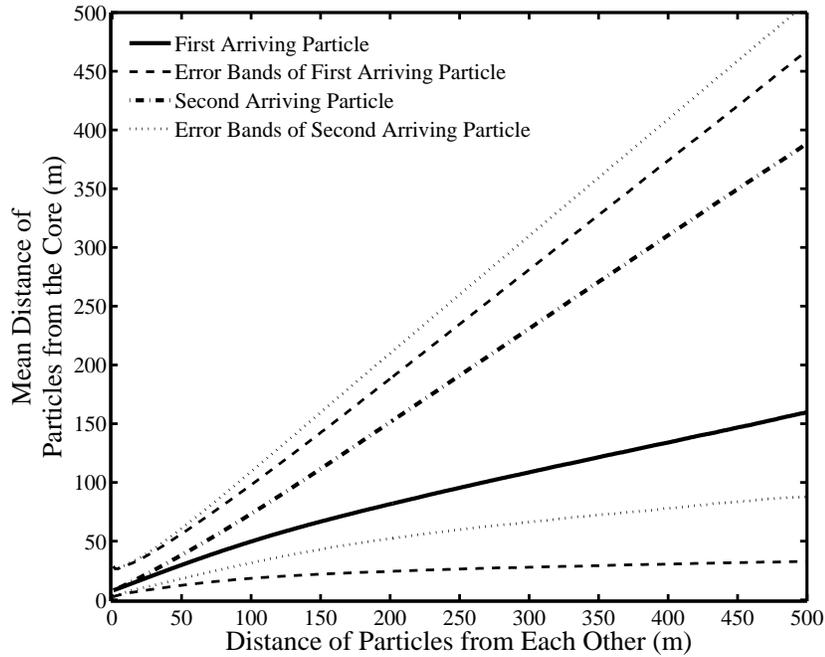}
\end{center}
\caption{Average distance of the first arriving particle (solid
curve) and the second arriving particle (dash-dot curve) from the
core versus their distance from each other for all secondary charged
particles of the showers. The related error bands are also depicted.
The results have been reached by additional 10,000 vertical single
energy (100 TeV) showers generated by CORSIKA. The low energy
hadronic model is FLUKA \cite{Fasso,Ferrari} and the high energy
hadronic model is QGSJETII \cite{Ostapchenko,Ostapch}.}
\label{AvDisBetFirSec}
\end{figure}

\begin{figure}
\begin{center}
\includegraphics*[width=1\textwidth]{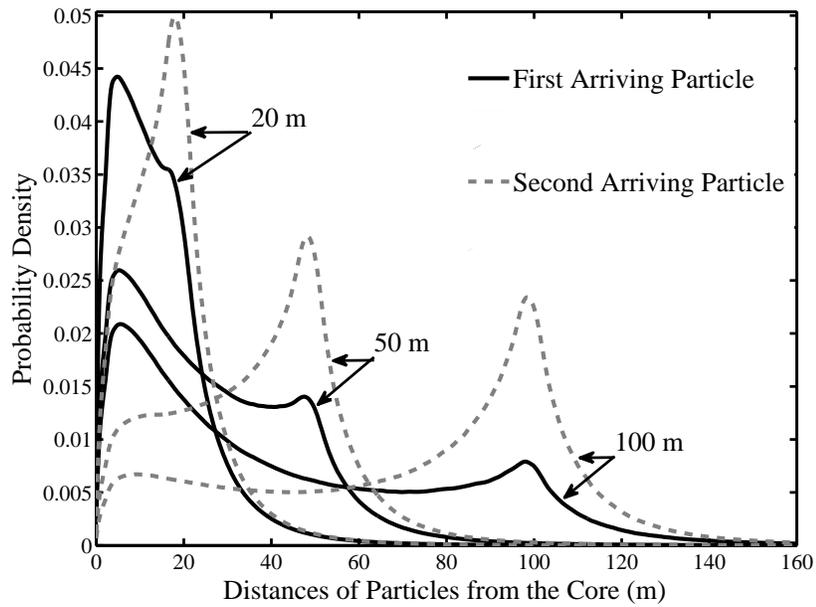}
\end{center}
\caption{Probability density of distances of two typical particles
of those EASs which were used in Fig. \ref{AvDisBetFirSec}. Three
groups of particle couples were selected with separations of 20 m
$\pm$ 0.5 m, 50 m $\pm$ 0.5 m and 100 m $\pm$ 0.5 m. Solid lines
belong to the first arriving particles and dashed lines belong to
the second arriving particles. In all three cases the first arriving
particle is more probable to be closer to the
core.}\label{ProDisFirSec}
\end{figure}

\begin{figure}[htp]
\begin{center}
\subfigure{\label{AveArrTim_1}\includegraphics[scale=0.4]{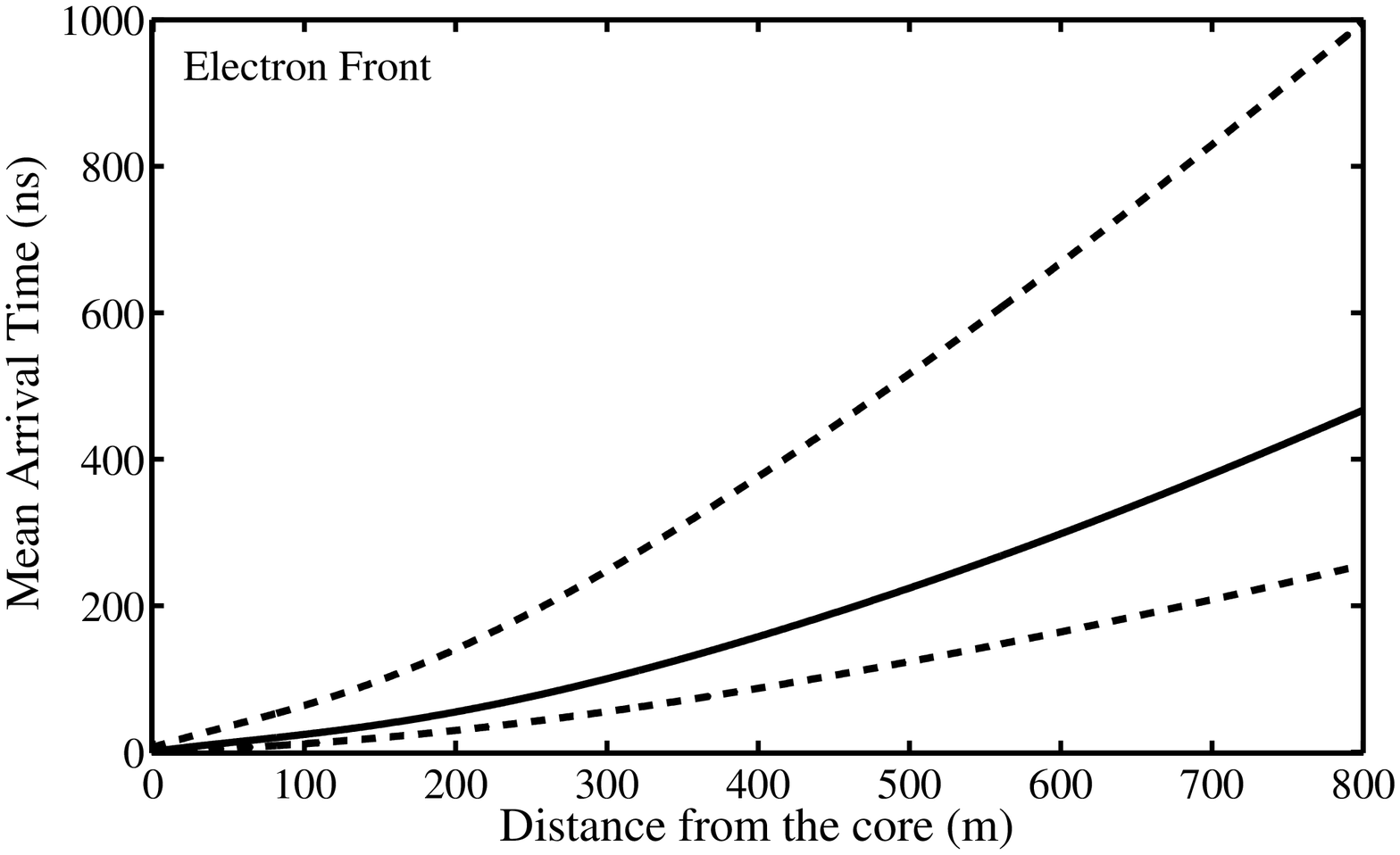}}
\subfigure{\label{AveArrTim_2}\includegraphics[scale=0.4]{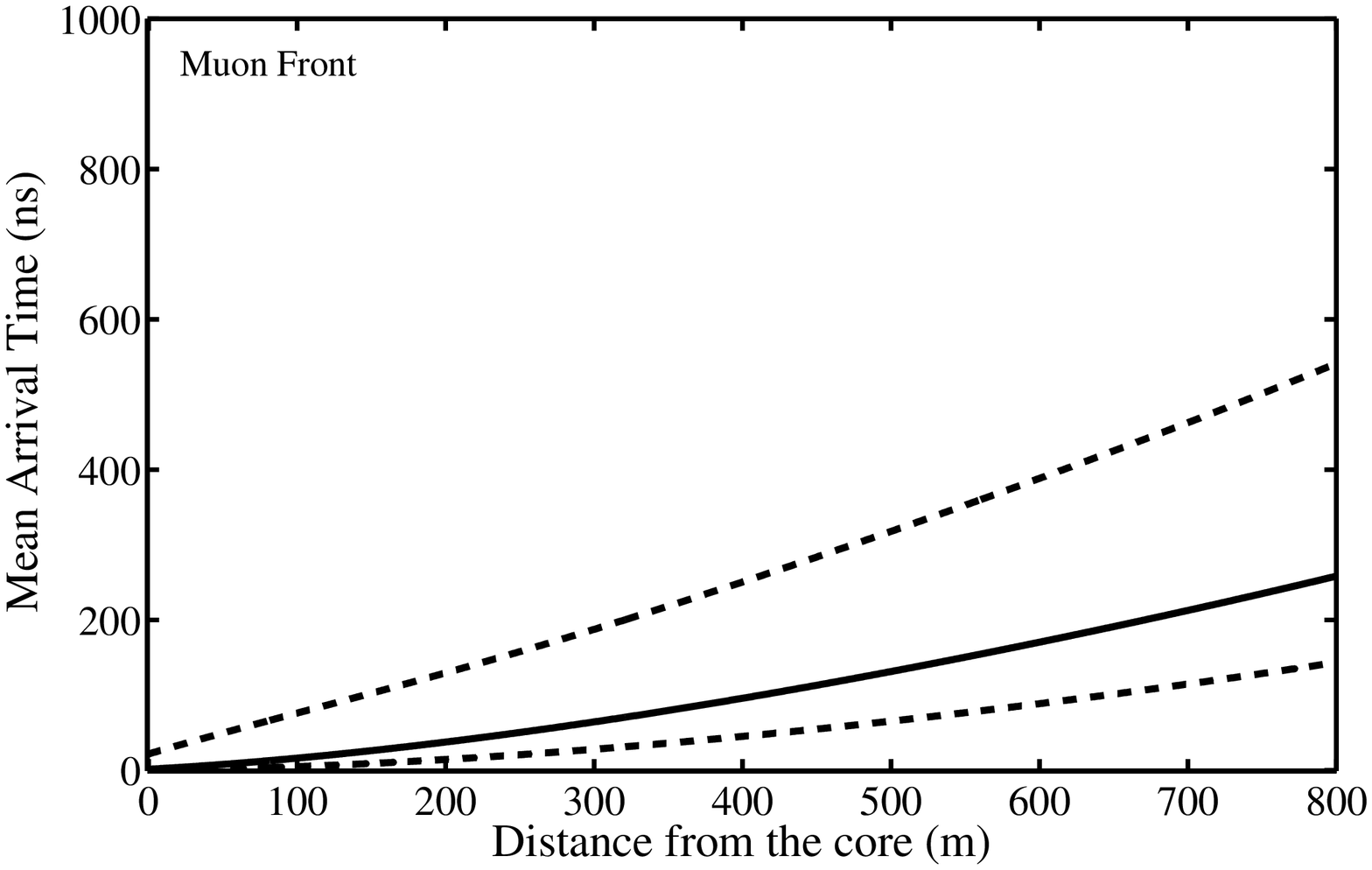}}
\subfigure{\label{AveArrTim_2}\includegraphics[scale=0.4]{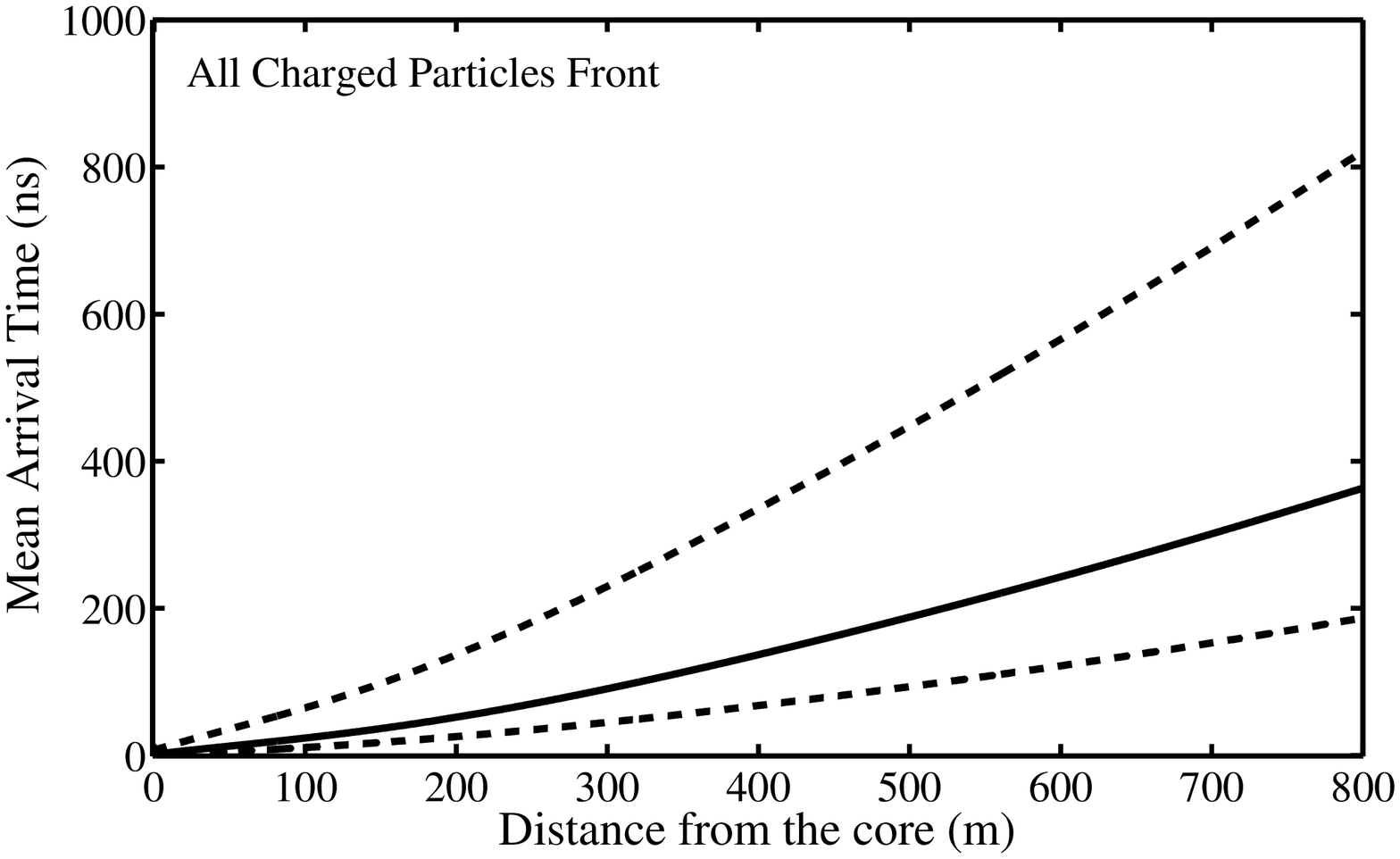}}
\end{center}
\caption{Mean arrival times of particles versus the distance from
the core for the same showers which were used in Fig.
\ref{AvDisBetFirSec}. Dashed lines are error bands.}
\label{AveArrTim}
\end{figure}

\begin{figure}
\begin{center}
\includegraphics*[width=1\textwidth]{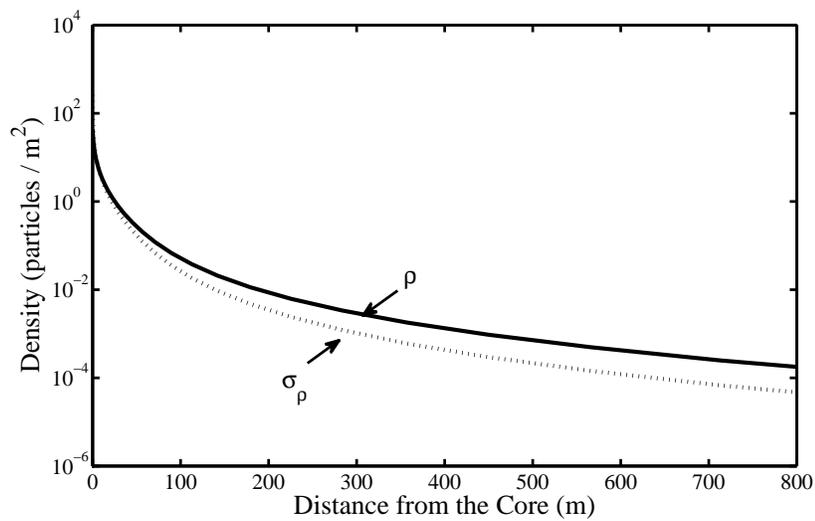}
\end{center}
\caption{The lateral density of charged particles, $\rho$, and its
standard deviation, $\sigma_\rho$, versus the distance from the core
for the VEASs used in Fig. 1.}\label{Density}
\end{figure}

An important feature of EASs is their spherical front, which is
approximately a spherical cap. Thus the particles in the core region
of a Vertical EAS (VEAS) reach to the ground level sooner than in
other regions. This feature can be demonstrated by considering the
fact that if we randomly select any two secondary particles of a
VEAS, the first particle reaching ground level on average is closer
to the core location. Fig. \ref{AvDisBetFirSec} shows that by
increasing the distance between two particles, the average distance
of the first arriving particle from the core increases slowly, while
the average distance of the second particle from the core increases
rather rapidly.\\
Another feature of EASs is that the particle density in the core
neighborhood is larger than in other regions. The smaller the
distance between two particles, the smaller their distance to the
core on average, as also follows from Fig. \ref{AvDisBetFirSec}.
Error bands also have been depicted in order to show the
fluctuations. They have been derived by calculating the RMS for
positive and negative errors separately. As can be seen the error
bands are asymmetric. Fig. \ref{ProDisFirSec} shows the reason of
this asymmetry.\\
The peak structures in Fig. \ref{ProDisFirSec} show that the first
arriving particle to the ground level is more likely to be closer to
the core location. For the small separations of particles, both
probability density functions have a unique pronounced peak close to
the core region. So, when the separation of two particles is small,
it is quite probable that both particles to be in approximately
equal distance from the core in opposite sides. For larger
separations, both probability densities have two peaks, one of which
is higher than the other. For the first arriving particle, the
higher maximum is near the core region, but the other one has the
same distance from the core as the distance between two particles.
For the second particle the situation is vice versa. So, when their
distances increase, although there is a small probability that the
first arriving particle will be found farther from the core than the
second one, it is more probable that the first arriving particle to
be much closer to the core region than the second arriving
particle.\\
The density of secondary particles quickly decreases by increasing
the distance from the core in contrast to the slow increase of the
average arrival time of particles shown in Fig. \ref{AveArrTim}. A
rough investigation shows that random fluctuation of the density of
the particles is much less than their arrival time fluctuations,
especially when they are far from the core. Fig. \ref{Density} shows
that the fluctuation of lateral density of the particles decreases
with the core distance rapidly in contrast to the arrival time
fluctuation which increases with the core distance (Fig.
\ref{AveArrTim}). So if we want to choose between particle density
information and arrival time information as a measure of the core
location, the particle density information is a more decisive
factor.
\section{A Method for Finding Core Location}
A simple approximate method to find the core position of an EAS is
to calculate the Center of Gravity (CG) of the Triggered Detectors
(TDs) which is a relatively good approximation for those EASs whose
cores are very close to the center of the array. We have introduced
a procedure which is capable to increase the precision of CG for
finding the approximate location of the core even when the core
position is close to the border of the array.\\
At first, we assume that every detector can just detect the first
crossing particle and is unable to detect any other particle during
an event. For simplicity, we investigated VEAS events at first and
then generalized the results to the inclined EASs.\\
Assume that we have positions and trigger times of all TDs of an
array during a VEAS event. TDs are indexed based on their trigger
times, starting with $i=1$ for the first TD. Fig. \ref{AveArrTim}
suggests that if $i<j<k$ then $\langle r_i\rangle<\langle
r_j\rangle<\langle r_k\rangle$, where $\langle r_i\rangle$ is the
average distance of the $i$th TD to the shower core. Therefore, by
taking into account Fig. \ref{AvDisBetFirSec}, this result can be
achieved: $\langle d_{ij}\rangle<\langle d_{ik}\rangle$ (where
$d_{ij}=|\vec{r}_i-\vec{r}_j|$). If we want to find the nearest
detector to the core, selecting the $i$th detector seems to be
logical, but because of timing fluctuations, a better choice will be
obtained by the following procedure: At first, we find the minimum
value of $d_{ij}$, $d_{ik}$ and $d_{jk}$. If $d_{jk}$ is the
smallest value then the $j$th detector is most likely nearest to the
core, since $j<k$.
\subsection{SIMEFIC Algorithm\label{SIMEFICSec}}
Based on the principles explained above, we developed the SIMEFIC
(SIeving MEthod for FInding Core) algorithm for eliminating
detectors far from the core. Let us assume that there are $N$ TDs in
an array during a VEAS. We form a matrix $D_{N\times N}$, whose
elements are $d_{ij}$s. In view of the fact that $D$ is a symmetric
matrix, we just consider the upper triangle of the matrix without
principal diagonal elements, which are zero.\\
\begin{linenomath}
\begin{equation*}
 D=\left(
    \begin{array}{ccccc}
      \times & d_{12} & d_{13} & \cdots & d_{1N} \\
      \times & \times & d_{23} & \cdots & d_{2N} \\
      \times & \times & \times & \cdots & d_{3N} \\
      \vdots & \vdots & \vdots & \ddots & \vdots \\
      \times & \times & \times & \cdots & \times \\
    \end{array}
  \right)
\end{equation*}
\end{linenomath}
Then we can find the smallest element ($d_<$) under the following
two conditions:
\begin{description}
  \item If $d_{ij}=d_{kl}$ and $i<k$ then $d_<=d_{ij}$
  \item If $d_{ij}=d_{ik}$ and $j<k$ then $d_<=d_{ij}$
\end{description}
Now we select the $i$th TD as the first detector of our near-core
list (because of its trigger time). Next, in the $i$th row, we find
the biggest element ($d_>$) under the following condition:
\begin{description}
  \item If all $d_{ij}$s ($j=i+1,...,N$) are different, the
  biggest element, e.g. $d_{ik}$, is $d_>$, and if $d_{ij}$
  and $d_{ik}$ are both the biggest elements ($d_{ij}=d_{ik}$),
  and $j<k$ then $d_>=d_{ik}$.
\end{description}
We now eliminate the $k$th TD as an off-core detector and remove the
$i$th and $k$th rows and columns of the matrix $D$ . By repeating
this procedure for the reduced matrix ($D_{N-2\times N-2}$), we will
reach a position in which half of the TDs are retained and half of
them are eliminated. Now, it is expected that the CG of the retained
TDs is a good measure of the core location.
\subsubsection{Inclined showers}\label{InSh}
Up to now, our discussion was limited to the VEASs. To generalize
this method to inclined EASs, the detector coordinates, and also the
trigger times need to be transformed into the coordinate system of
the shower. The coordinates of the detectors on the ground level are
$x_i$, $y_i$, and $z_i=0$. Now, these coordinates are transformed to
a new coordinate system whose $x$ and $y$ axes are perpendicular to
the arrival direction. To perform this, we select an arbitrary point
(e.g. the CG of all detectors) as the origin of coordinate system.
Then the coordinates of the detectors are first rotated
counterclockwise around the $z$ axis by the angle $\phi$ and next
the new coordinates are rotated clockwise around the $y^{'}$ by the
angle $\theta$. The $z$ components in the final coordinate system
are $z_i^{''}=-x_i\sin\theta\cos\phi-y_i\sin\theta\sin\phi$.
Accordingly the trigger times of the detectors are transformed to
$t_i^{''}=t_i+z_i^{''}/c$ where c is the speed of light. Now the
inclined showers are treated as the vertical ones.
\subsubsection{Arrays of DSDs}
For arrays with DSDs the algorithm is extended as follows. Another
matrix, $N$, is formed, the elements of which are $n_{ij}=n_i+n_j$
($n_i$ is the number of the detected particles by $i$th detector) in
correspondence with $d_{ij}$ elements of the matrix $D$. Now, we
first select a pair of detectors with maximum $n_{ij}$ ($n_{max}$)
under the following two conditions:
\begin{description}
  \item If $n_{ij}=n_{kl}$ with $i<k$, and if $d_{ij}>d_{kl}$, then
$n_{max}=n_{kl}$ and if $d_{ij}\leq d_{kl}$, then $n_{max}=n_{ij}$.
  \item If $n_{ij}=n_{ik}$ with $j<k$, and if $d_{ij}>d_{ik}$, then
$n_{max}=n_{ik}$ and if $d_{ij}\leq d_{ik}$, then $n_{max}=n_{ij}$.
\end{description}
Assume that $n_{ij}$ has been selected as $n_{max}$ and $n_i<n_j$,
then $j$ is the near core detector (if $n_i>n_j$, $i$th detector
will be chosen). If $n_i=n_j$, the trigger times of the two
detectors $i$ and $j$ will be the determining factor for choosing
between $i$ and $j$ as the near core detector. Other stages of the
algorithm are the same as before.
\subsection{Reconstruction of a Shower Geometry}
As the precision of final results of SIMEFIC algorithm is tied to
the precision of arrival direction measurement, a method has been
introduced here that can be used for more precise arrival direction
reconstruction.\\
Assume that we have $\{P_i\}$ set of the locations and trigger times
of all TDs at the beginning of the calculation. By using PWFA and
SIMEFIC algorithm on the set $\{P_i\}$, a subset of it,
$\{P_i^{'}\}$, will be found whose members are half of the first
set. Now, PWFA is used in shower direction reconstruction of the
$\{P_i^{'}\}$ set from which the arrival direction is calculated
with more precision. The reason for higher precision, of this
technique in comparison with the PWFA used for the set $\{P_i\}$, is
that arrival times have less fluctuation in near core region than
those in far regions from the core. Furthermore, the front of an EAS
is smoother in near core region than that in far regions. In view of
these facts, SIMEFIC method presents a better approximation for
arrival direction in comparison with the PWFA, which uses data of
all TDs. Now, we propose a refinement to the SIMEFIC method by
repetition of the above procedure:
\begin{enumerate}
  \item Reconstruct arrival direction by using $\{P_i^{'}\}$ set with PWFA,
   $(\theta^{'}, \phi^{'})$.
  \item Use SIMEFIC algorithm for the original set, $\{P_i\}$
   (not on set $\{P_i^{'}\}$ which has $N/2$
   members) and arrival direction $(\theta^{'}, \phi^{'})$,
   to obtain another set $\{P_i^{''}\}$.
  \item Replace $\{P_i^{'}\}$ by $\{P_i^{''}\}$ and repeat the procedure.
\end{enumerate}
This procedure is repeated several times using the original set to
refine the arrival direction. If two successive runs yield
approximately the same results, the process will be terminated. In
our simulation, repeating for 3 times was enough.
\section{Simulations with CORSIKA}\label{sim_sec}
The geographical coordinates of the prototype of Alborz observatory
location (in Tehran, $35^{\circ}$N, $51^{\circ}$E and 1200 m above
sea level) have been imposed in CORSIKA EAS simulations. Geomagnetic
field components of Tehran are $B_x=28.1 \mu$T and $B_z=38.4 \mu$T.
For the simulation of low energy hadronic interactions, GHEISHA
\cite{Fesefeldt} package, and for high energy cases QGSJET01 package
have been used. Zenith angles of primary particles were chosen
between $0^{\circ}$ and $60^{\circ}$. The compositions of primary
particles have been $90\%$ protons and $10\%$ helium nuclei. This
ratio was assumed to be constant over entire energy range. The
energy of the primary particles ranges from 100 TeV to 5 PeV. Other
parameters are CORSIKA default values (e.g. default spectrum
index of $-2.7$).\\
The assumed array is a square with an area of $200\times200$
m$^{2}$, composed of $1\times1$ m$^{2}$ detectors on a square
lattice with a $5$ m lattice constant (total of $41\times41$
detectors). Ground arrays are commonly made up of scintillation
detectors which are not sensitive to the position of the crossing
particles and are not able to exactly measure how many particles
have passed through them. Therefore, the following assumptions have
been applied for the
simulation.\\
When the first arriving particle passes through a detector, its
arrival time is considered as the trigger time of that detector.
Obviously, the coordinates of this first crossing particle must be
within $\pm0.5$ m of the center of the detector in order to consider
the detector as a TD. The next arriving particles which pass through
this hypothetical detector are not taken into account.\\
The CG of TDs for the array whose detectors are DSDs will be
considered as follows:
\begin{linenomath}
\begin{align*}
x_{CG}&=1/N\displaystyle\sum\limits_{i}n_ix_i,\\
y_{CG}&=1/N\displaystyle\sum\limits_{i}n_iy_i
\end{align*}
\end{linenomath}
where $n_i$ is the pulse height (the number of crossing particles)
of each detector and $N=\displaystyle\sum\limits_{i}n_i$.
For non DSDs, we set $n_i=1$ for all TDs.\\
If the coordinates of the EAS true core are denoted by $(x_{tc},
y_{tc})$, the distance between the true core and the CG will be:
\begin{linenomath}
\begin{equation*}
    r=\sqrt{(x_{CG}^{''}-x_{tc}^{''})^{2}+(y_{CG}^{''}-y_{tc}^{''})^{2}}
\end{equation*}
\end{linenomath}
where $x^{''}, y^{''}$ are coordinates in the rotated coordinate
system introduced in sec. \ref{InSh}.\\
\subsection{Results}
\begin{figure}
\begin{center}
\includegraphics*[width=1\textwidth]{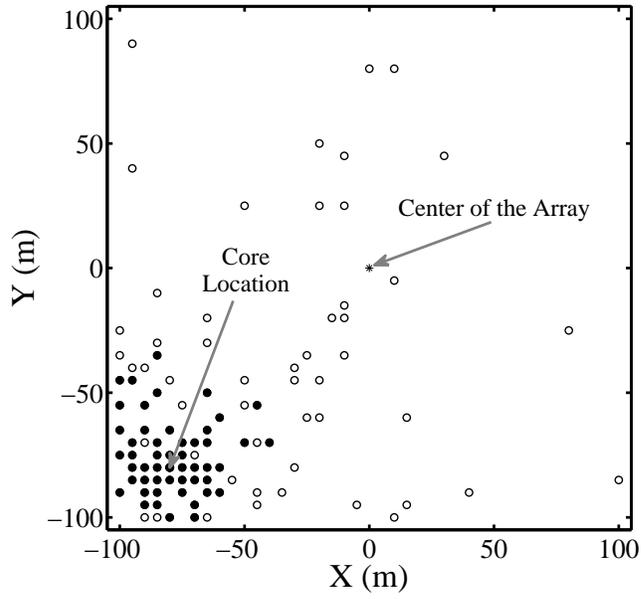}
\end{center}
\caption{The TDs which are accepted by SIMEFIC method are shown by
the bold circles and the omitted ones by the empty circles. The
position of the true core and the center of the array have been
shown. The zenith and azimuth angles of this shower are
$29.5^{\circ}$ and $329.5^{\circ}$ respectively, and its energy is
157.5 TeV.} \label{SampleShower}
\end{figure}

\begin{figure}[htp]
\begin{center}
\includegraphics*[width=1\textwidth]{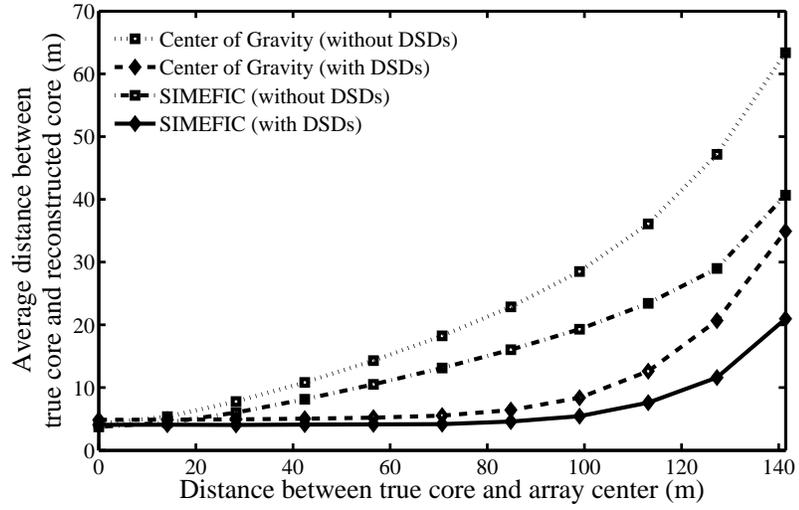}
\end{center}
\caption{The comparison of the CG of all TDs and those which have
been selected by SIMEFIC. The size of error bars for each data point
is less than the size of the symbols used for them.} \label{ComCGSI}
\end{figure}

\begin{figure}
\begin{center}
\includegraphics*[width=1\textwidth]{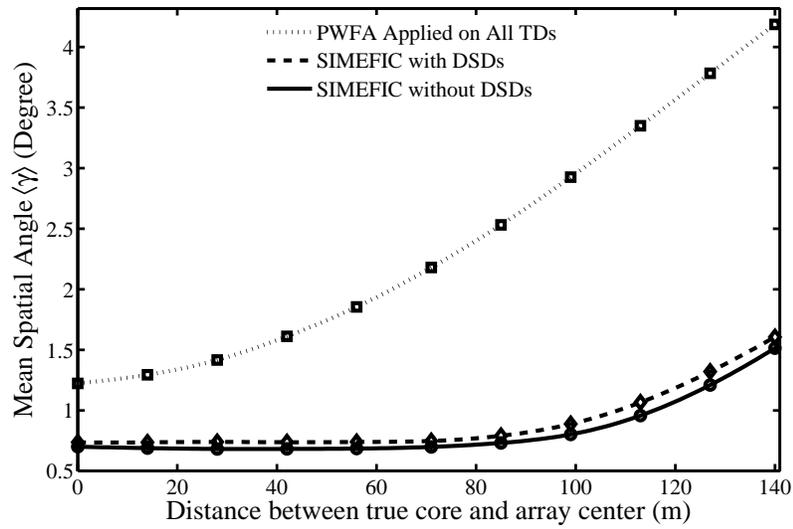}
\end{center}
\caption{Mean value of angle between the EAS primary direction and
the direction which has been found by SIMEFIC algorithm assuming the
array with non DSDs (solid curve), SIMEFIC algorithm assuming the
array with DSDs (dashed curve) and by PWFA applied on all TDs
(dotted curve) versus the distance of the true core from array
center.} \label{AngleComp}
\end{figure}

Fig. \ref{SampleShower} shows an EAS which its true core is on point
$(-80, -80)$. Detectors accepted by SIMEFIC method are shown by the
bold circles and the omitted ones by the empty circles. It is clear
that even when the true core is near the edge of the array, this
algorithm has approximately chosen proper detectors, while, some of
the detectors that are far from the core but near to each other have
not been chosen. This is an important effect of using the timing
information. This method is precise up to the point that almost half
of the TDs are symmetrically spread around the core.\\
In Fig. \ref{ComCGSI} and Fig. \ref{AngleComp}, we consider the true
core on the diagonal line on the points $(10i, 10i)$ ($i=1,...,10$),
with the center of the array at point $(0, 0)$. Trigger condition
for the EASs, whose specifications have been introduced at the first
part of the current section, was triggering at least 68 detectors
out of 1681 detectors by secondary particles. 3000 of the accepted
EASs have been averaged for each data point shown in Fig.
\ref{ComCGSI} and Fig. \ref{AngleComp} and since the size of error
bars for each data point is less than the size of the symbols used
for them, the error bars are omitted in these figures.\\
Fig. \ref{ComCGSI} shows the results of using SIMEFIC method and the
CG of all TDs for finding core locations. It is clear that the CG of
all TDs is only precise for those EASs whose core are near the
central part of the array and, by increasing the distance of EAS
core from the center of the array, the error increases gradually.
But, in SIMEFIC method, the precision, even up to 50\% of the length
of the array, is approximately the same. As we expect, the results
for an array with DSDs are better than those for the array without
DSDs.\\
Fig. \ref{AngleComp} shows the angle between the EAS primary
direction provided by CORSIKA and the direction which has been found
by the SIMEFIC method and also by PWFA (applied on all TDs) versus
the distance of true core from array center. It is clear that
SIMEFIC algorithm has significantly improved PWFA. Furthermore, Fig.
\ref{AngleComp} shows that the results obtained with non DSDs are
slightly more precise than those of DSDs. We think that the reason
behind this unexpected result is that for finding direction of an
EAS, it is better to give all of the detectors the same weight and
choose them with the same probability around the true core location.
When we discard density information, all of the TDs have the same
probability to being chosen, so the algorithm will select TDs
symmetrically around the core. Due to the stochastic nature of
secondary particles of the shower, the symmetric selection will be
rather destroyed with DSDs and those TDs which have more detected
particles have a more chance for selection.
\section{Conclusion}
In this investigation, we have developed a simple method for finding
the position of an EAS core, and applied PWFA to an optimized data
set for finding the EAS arrival direction. In this method, we have
used the information of positions of the TDs on the ground level as
well as the trigger times. In this method, distances between all
pairs of TDs are measured and then TDs with minimum separation and
smallest trigger time are chosen, while TDs with the maximum
separation and largest trigger times are omitted. Finally, the CG of
the chosen TDs is used for finding core location of EAS.\\
The essence and scheme of the method envisioned by examining
vertical showers have been also formulated for inclined showers. An
operational algorithm were developed and tested over $3\times10^5$
simulated EAS events generated by CORSIKA package. The proposed
analysis technique is adequate for simple EAS arrays without DSDs,
though it has been generalized for arrays with DSDs, and its results
for finding core location are more accurate with DSDs.\\
The precision of finding EAS core location is improved to about the
array's lattice constant for an EAS whose core falls within the
central region of the array and to about 4 times lattice constant
for those falling close to the edge of the array. An improvement of
about 58\% has been reached in comparison with the CG of all TDs for
the above mentioned geometry and configuration of the array.
Furthermore, by using the PWFA for the TDs selected by this method,
the angular resolution of the primary arrival direction is improved
significantly compared to the case of using PWFA on all TDs of the
array during an event. The angular error ranging from $1.2^\circ$ to
$4.2^\circ$ for the case of using all TDs is reduced to values
ranging from $0.4^\circ$ to $1.5^\circ$ corresponding to showers
falling near the center or on the edge of the array. Again, this is
on average an improvement of about $89\%$ in comparison with the
simple PWFA.\\
It should be noted that these improvements belong to the results of
our simulations for the assumed array of 1681 normal detectors.
Obviously, the improvement depends on the size of EAS array, its
spacing, and its detectors type. Further improvements are expected
by optimally combining the information contained in the two
matrices, $D$ and $N$ mentioned in sec. \ref{SIMEFICSec}. In our
future attempts we will investigate simulated data to find the
optimal method of combining the information of these two matrices
($D$ and $N$). The method for finding core location could be used as
an improved first guess in order to seed the common fitting
methods.\\
%\end{spacing}
\section*{Acknowledgements}
This research was supported by a grant from the office of vice
president for science, research and technology of Sharif University
of Technology. The authors are very grateful for the invaluable and
constructive comments of the anonymous referee.

%\begin{references}

\clearpage

\clearpage

\clearpage

\clearpage

\clearpage

\clearpage

\clearpage

\end{document}